\documentclass[sigconf]{acmart}

\AtBeginDocument{%
  \providecommand\BibTeX{{%
    \normalfont B\kern-0.5em{\scshape i\kern-0.25em b}\kern-0.8em\TeX}}}
\usepackage{bm}

\usepackage{pbox}

\newcommand{\mybox}[1]
{
\vspace{1.8mm}
\noindent \hspace{-1mm} 
\setlength\fboxsep{1mm}
\fbox{\parbox{\dimexpr\linewidth-2\fboxsep-2\fboxrule}{\itshape #1}}
}

\setcopyright{acmcopyright}
\copyrightyear{2024}
\acmYear{2024}
\acmDOI{-}

\acmConference[Conference'24]{EDBT '24: International Conference on Extending Database Technology}{2024}{}
\acmBooktitle{}
\acmISBN{978-1-4503-XXXX-X/18/06}

\usepackage{algorithmic}
\usepackage{graphicx}
\usepackage{textcomp}
\usepackage{xcolor}
\usepackage[linesnumbered]{algorithm2e}
\usepackage{amsthm}

\usepackage{graphicx}

\usepackage{balance}
\usepackage{array}
\usepackage{tabulary}
\newcolumntype{K}[1]{>{\centering\arraybackslash}p{#1}}

\usepackage{diagbox}
\usepackage{caption}

\usepackage{relsize}
\usepackage{verbatim}

\usepackage{amsthm}
\usepackage{hyperref}
\usepackage{subcaption}
\usepackage{enumitem}
\usepackage{booktabs} % For formal tables
\usepackage{multirow}
\usepackage{tabularx}

\usepackage[para]{footmisc}

\usepackage{tikz}
\usetikzlibrary{trees}
\usetikzlibrary{positioning}
\usetikzlibrary{matrix}
\definecolor{myyellow}{RGB}{250, 246, 145}
\definecolor{myorange}{RGB}{245, 203, 130}
\definecolor{mygreen}{RGB}{211, 247, 188}
\definecolor{myblue}{RGB}{86, 180, 233}
\def\BibTeX{{\rm B\kern-.05em{\sc i\kern-.025em b}\kern-.08em
    T\kern-.1667em\lower.7ex\hbox{E}\kern-.125emX}}

\newtheorem*{definition}{Definition}

\newcounter{AsteriosNOC}
\stepcounter{AsteriosNOC}

\newcounter{ChristosNOC}
\stepcounter{ChristosNOC}

\newcounter{RihanNOC}
\stepcounter{RihanNOC}

\newcounter{MariosNOC}
\stepcounter{MariosNOC}

\newcounter{GeorgeNOC}
\stepcounter{GeorgeNOC}

\newcommand{\para}[1]{\vspace{1.2mm}\noindent\textbf{#1.}}

\newcommand{\paranofull}[1]{\vspace{1.2mm}\noindent\textbf{#1}}

\newcommand{\parait}[1]{\vspace{1.2mm}\noindent\emph{#1.}}

\usepackage{hyperref}

\usepackage{graphicx}

\newcommand{\method}[0]{SiMa}

\usepackage{multirow}
\usepackage[linesnumbered]{algorithm2e}
\usepackage{amsthm}

\theoremstyle{definition}
\newtheorem{thm}{Theorem}[section]
\newtheorem{problem}[thm]{Problem}
\usepackage{xcolor}
\definecolor{commentgreen}{rgb}{0,0.5,0}

\SetCommentSty{mycommfont}
\begin{document}

% \newtheorem{problem}{Problem}

\title{SiMa: Effective and Efficient Matching Across Data Silos Using Graph Neural Networks}

\author{Christos Koutras \quad Rihan Hai \quad Kyriakos Psarakis \quad Marios Fragkoulis \quad Asterios Katsifodimos}
\affiliation{%
  \institution{Delft University of Technology}
  \streetaddress{P.O. Box 1212}
  \postcode{43017-6221}
  \country{\vspace{-.5mm}}
}
\email{initial.lastname@tudelft.nl}

\renewcommand{\authors}{Christos Koutras, Rihan Hai, Kyriakos Psarakis, Marios Fragkoulis, Asterios Katsifodimos}

\begin{abstract}
How can we leverage existing column relationships within silos, to predict similar ones across silos? Can we do this efficiently and effectively? Existing matching approaches do not exploit prior knowledge, relying on prohibitively expensive similarity computations.
In this paper we present the first technique for matching columns across data silos, called \method, which leverages Graph Neural Networks (GNNs) to learn from existing column relationships within data silos, and dataset-specific profiles. The main novelty of \method{} is its ability to be trained incrementally on column relationships within each silo individually, without requiring the consolidation of all datasets in a single place. Our experiments show that \method{} is more effective than the -- otherwise inapplicable to the setting of silos -- state-of-the-art matching methods, while requiring orders of magnitude less computational resources. Moreover, we demonstrate that \method{} considerably outperforms other state-of-the-art column representation learning methods.
\end{abstract}

\maketitle

\section{Introduction}
\label{sec:intro}
Given a large set of datasets spread across different data silos \cite{mansour2022federated}, as well as example column relationships within those silos, how can we detect pairs of dataset columns, that are joinable or unionable across silos? Can we do this \emph{efficiently} and \emph{effectively}?
\begin{figure}[t]
    \centering
    \includegraphics[width=.99\columnwidth]{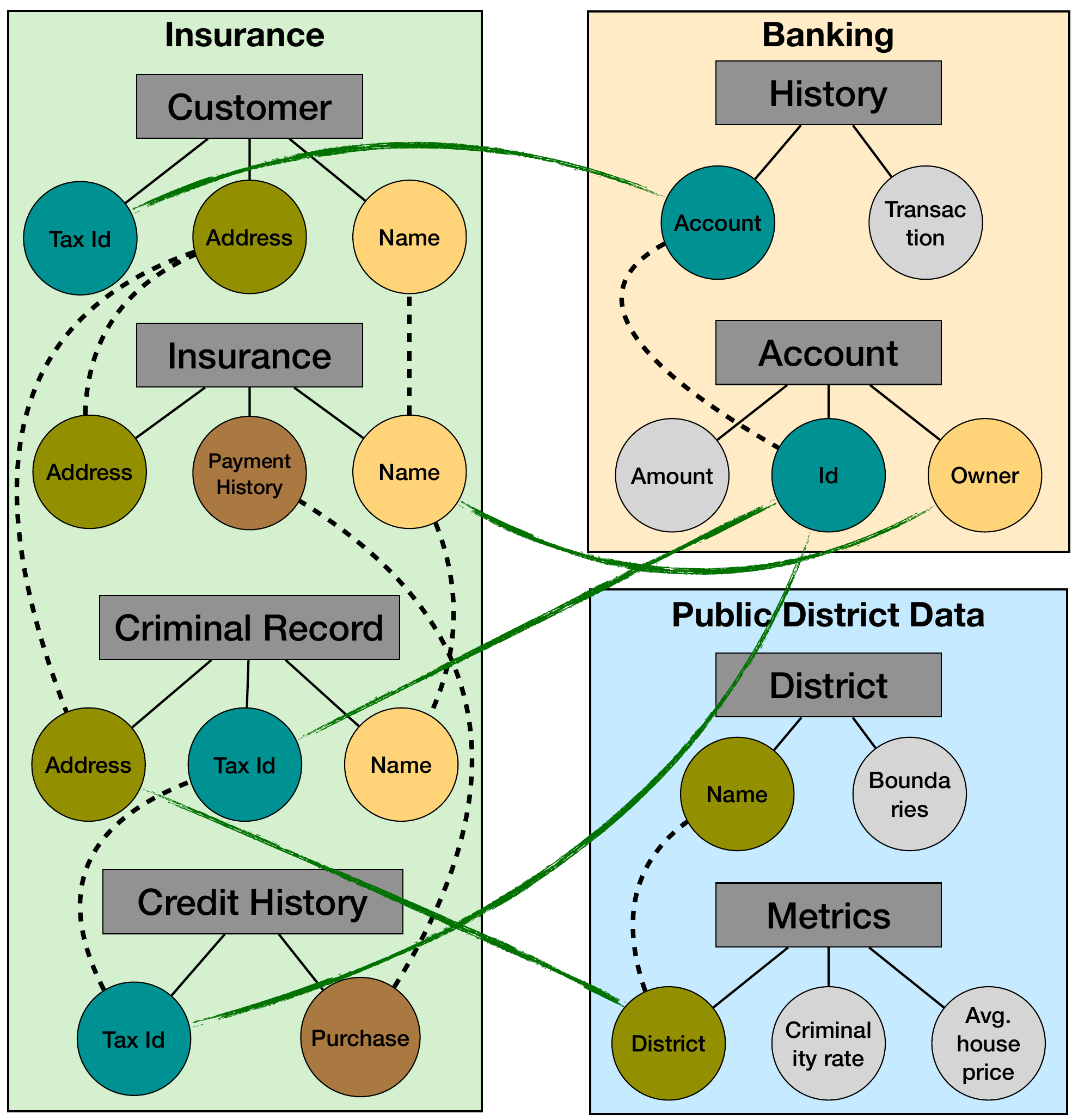}
    \caption{Three typical data silos in the banking industry.}
    \label{fig:silos}
    % \vspace{-3mm}
\end{figure}
Organizations nowadays accumulate large numbers of heterogeneous datasets in data lakes, with the goal of gaining insights by combining those datasets. The structure (e.g., departments, teams, locations) of organizations,  but also the sheer scale of their data lakes, force organizations to establish barriers for their data assets, leading to the phenomenon of \emph{data silos}: disjoint and isolated collections of datasets, belonging to different stakeholders. Interestingly, data silos may even exist within the same organization, as individual teams enforce their own conventions and formats, as well as encapsulate knowledge about their data assets. Silo-ing data impedes collaboration and information sharing among different groups of interest.

\para{Running Example} Consider an organization in the banking industry as depicted in \autoref{fig:silos}. Employees of the banking silo already know the relationships between their datasets (black dotted lines), i.e. columns from tables inside the silo that are semantically related (storing values that refer to the same semantic type). However, the possible relationships between the banking silo and the other two silos (green lines) are missing, i.e. columns of the same semantic type,  residing in different silos. Data scientists building ML models can benefit from dataset augmentation in terms of extra data points (by finding other unionable datasets) and/or extra features (by finding other joinable datasets) from other data silos \cite{chepurko2020arda}.

\para{Column Matches Within Silos} To enable collaboration across departments and teams, organizations build and maintain dataset metadata catalogs \cite{fernandez2018aurum,halevy2016goods}: a graph structure that encapsulates relationships among datasets. Typically, \emph{within} a given silo, one can enrich a metadata catalog with PK-FK relationships using schema information and automated data profiling techniques \cite{abedjan2015profiling} as well as joinability/unionability relationships, using matching techniques\cite{koutras2021valentine}. Moreover, such relationships can be derived from domain experts, query logs \cite{nandi2009hamster, bharadwaj2021discovering}, and even data science notebooks \cite{psallidas2022data}. However, discovering relationships among columns \emph{across} data silos is very challenging \cite{mansour2021federated}.

\begin{figure*}[t!]
    \centering
    \includegraphics[width=2.12\columnwidth]{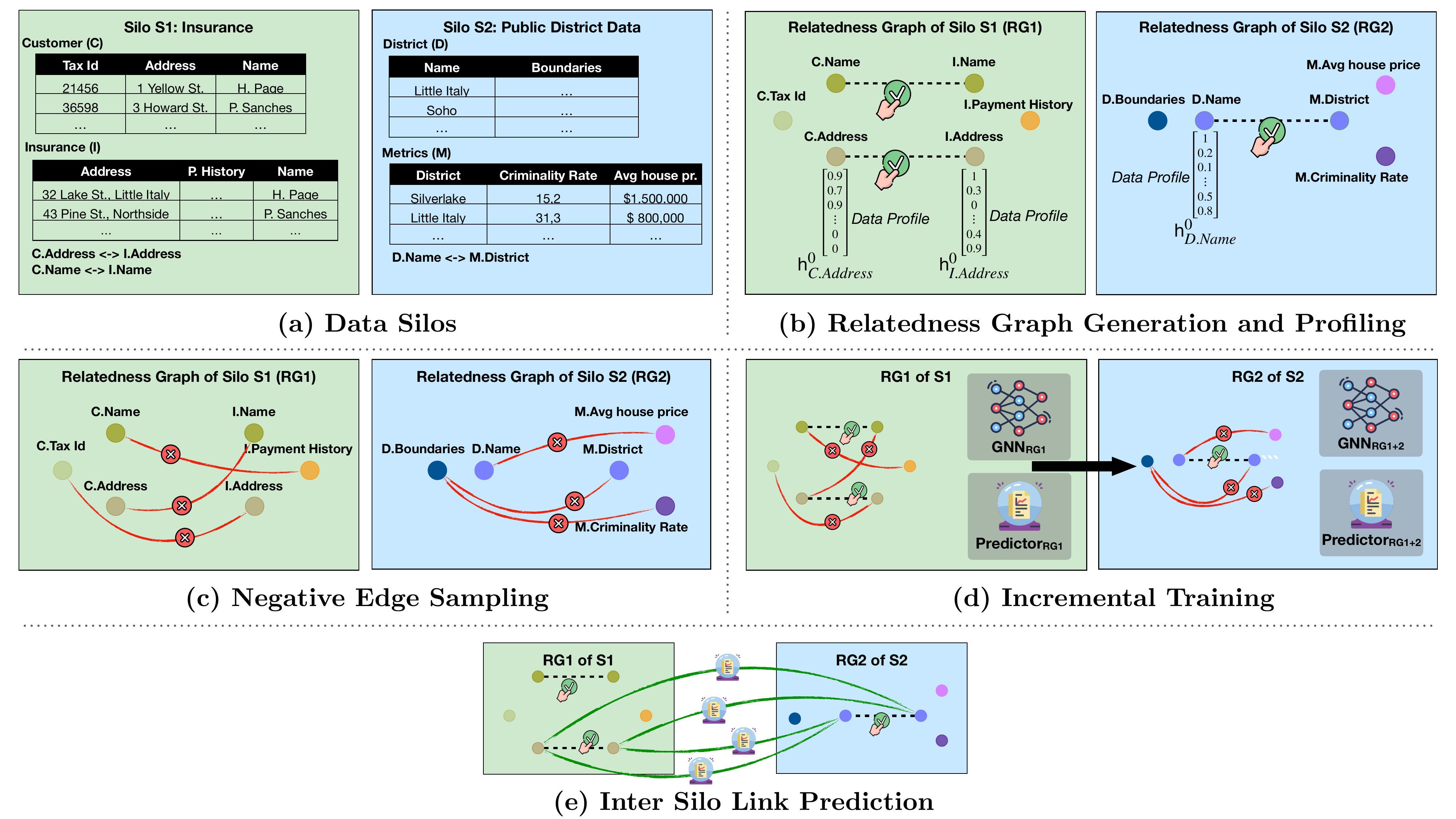}
    \vspace{-6mm}
    \caption{SiMa overview: (a) depicts data silos and their column matches which are transformed into relatedness graphs ((b)-\autoref{sec:model}), where nodes represent columns and receive their initial features from a tabular data profiler (\autoref{sec:profile}). Then, negative edges are being sampled from each relatedness graph as shown in  (c) (\autoref{sec:sampling}) and a link prediction model is being trained based on an incremental training scheme depicted in (d) (\autoref{sec:incre_train}). Finally, using the trained model we are able to predict relationships among columns from  different silos as depicted in (e).}
    \label{fig:sima}
    \vspace{-4mm}
\end{figure*}

\para{Existing solutions} In the data management research, the problem of finding relationships among datasets has been investigated in three different contexts (more details in \autoref{sec:relatedwork}): \emph{i}) \textit{schema matching}, with a multitude of automated methods \cite{rahm2001survey,  zhang2011automatic, lehmberg2017stitching, fernandez2018seeping, koutras2020rema, cappuzzo2020creating}; $ii)$~\emph{related-dataset search} \cite{fernandez2018aurum, nargesian2018table, zhu2019josie,bogatu2020dataset, zhang2020finding, fan2022semantics, khatiwada2022santos, bogatu2022voyager}, and $iii)$~\emph{column-type detection} \cite{hulsebos2019sherlock, zhang13sato}. In short, traditional schema matching methods are \emph{a)} computationally and resource expensive; \emph{b)} they cannot always be employed in the setting of data silos as they require co-locating all datasets to calculate similarities; \emph{c)} they do not leverage existing knowledge within silos. Related-dataset search methods are not applicable to the matching problem as their goal is to  search top-k related datasets to a given dataset, sacrificing recall for precision. To tune such methods for discovering more column matches (increase the recall), we would need to set k to a large value, which could dramatically affect the quality of the results (high false positive rates, thus lower precision). Finally, column-type detection requires knowing the types of all columns in advance, alongside massive training data.

\para{\method{}: an efficient \& effective silo matcher\footnote{An early version of this work has been presented in a non-archival workshop \cite{simatrl}.}} In this paper we propose \method{}, a novel approach to the problem of discovering relationships between tabular columns across data silos (Figure~\ref{fig:sima}). \method{} is based on the observation that \textit{within} silos we can find existing matches among columns and train a ML model that learns to predict column relationships \textit{across} silos: \emph{i)}~equi-joinable, \emph{ii)}~fuzzily-joinable, \emph{iii)}~unionable columns of the same domain.

SiMa leverages the representational power of \emph{Graph Neural Networks} (GNNs). However, employing GNNs for the purposes of matching across data silos is far from straightforward, as we need to: \emph{i})~transform tabular data to information-preserving graphs, \emph{ii})~initialize nodes with suitable features, \emph{iii})~introduce non-trivial negative-sampling techniques and training schemes to optimize the learning process. SiMa provides with effective and efficient solutions to each of these problems, proceeding as shown in \autoref{fig:sima}. 

\noindent In short this paper makes the following contributions:
\vspace{-2mm}
\begin{itemize}
\item We define the problem of \emph{matching across data silos} (\S~\ref{sec:matching}). 
\item We propose \emph{\method}, a generic and inductive GNN-based learning framework, which discovers relatedness across different data silos. 
To the best of our knowledge, our work is the first to generalize local matches within a silo, to links across silos.

\item We show how to represent data silos, and the knowledge about matches among datasets inside those silos as graphs, turning the problem \emph{matching across data silos}  into a \emph{link prediction} task (\S~\ref{sec:cas}).

\item We propose two optimization techniques, \emph{negative edge sampling}  and   \emph{incremental model training}, which improve  the training efficiency and effectiveness of our GNN for the purposes of matching across silos (\S~\ref{sec:opt}).

\item With experiments (\S~\ref{sec:experiments}) over real-world data from several domains and open datasets, \method{} demonstrates significant effectiveness gains with orders of magnitude run-time performance savings (up to 600x) compared to traditional (and inapplicable) schema matching methods and column representation techniques.
% \vspace{2mm}
\end{itemize}

\vspace{1mm}
\noindent The datasets, ground truth and code of this work, are available at \url{https://github.com/delftdata/SiMa}.

\section{Approach Overview}
Five aspects comprise \method{}'s approach: a relatedness graph, data profiles, a learning method, a prediction method, and an optimization process.

\parait{-- Relatedness Graph (\autoref{sec:model})} As shown in \autoref{fig:sima}b, \method{} transforms each silo's set of columns and the respective matches among them (\autoref{fig:sima}a) into nodes and corresponding edges, thus creating as many graphs as data silos.

\parait{-- Data Profiles (\autoref{sec:profile})} For each column, \method{} builds a profile of 987 features (\autoref{fig:sima}b), such as the number of numerical values among the instances or character-level aggregates \cite{abedjan2015profiling,hulsebos2019sherlock}. These column profiles facilitate the training process of a Graph Neural Network (GNN). 

\parait{-- Learning from profiles and graph information (\autoref{sec:training})} Each data profile is used as a feature vector of each node in the relatedness graph (\autoref{fig:sima}b). Using GNNs, \method{} takes into consideration the profiles and the graph edges in order to learn how to incorporate the graph's neighborhood information together with the features of each node.

\parait{-- Predicting matches across silos (\autoref{sec:pipeline})} Finally, as depicted in \autoref{fig:sima}e, \method{} uses the learned graph embeddings from the GNNs to capture similarity among columns, and discover matches across data silos.  We do this by fine tuning a link prediction model, enabling \method{} to decide whether there could be a match between a pair of columns or not. 

\parait{-- Optimizing the GNN learning process (\autoref{sec:opt})} \method{} applies sophisticated negative edge sampling techniques on the graphs (\autoref{fig:sima}c) to fine tune the prediction ability of the GNNs  by  leveraging the knowledge inside each data silo (\autoref{sec:sampling}). Moreover, as shown in \autoref{fig:sima}d, with incremental training (\autoref{sec:incre_train}) \method{}  not only improves its match prediction ability, but also allows silos to train a GNN \textit{individually}, without having to consolidate all data profiles in one place.

\section{Related Work}
\label{sec:relatedwork}
The closest work to this paper is traditional schema matching methods that were not designed for the problem of matching across silos (\autoref{sec:rel-sm}), as well as dataset discovery \& semantic type detection that do not address our problem directly (\autoref{sec:rel-dd}).

\subsection{Schema Matching} 
\label{sec:rel-sm}
A natural choice to bridge data silos would be to employ  \emph{schema matching} \cite{rahm2001survey, gal2011uncertain}, namely a set of methods responsible for finding matches among elements of disparate datasets based on various similarity criteria (e.g. \emph{Jaccard similarity}). Schema matching is a well-studied research topic, with various methods mainly focusing on finding matches between pairs of tables \cite{do2002coma, madhavan2001generic, zhang2011automatic, chen2018biggorilla, shraga2020adnev, cappuzzo2020creating}. However, 
existing matching methods assume global access to all datasets so that they can compute similarities between pairs of columns. Across data silos, this is usually impossible, since the stakeholders are not willing to share data with each other \cite{miller2018open}.

\para{Statistics-based methods} One can base a matching method's similarity calculations on data statistics \cite{do2002coma, zhang2011automatic}: first compute statistics of columns within a silo, and carry those statistics over to other silos for similarity calculations. However, it is often the case that characteristics of values across silos can differ substantially, even for the same semantic types (e.g., names of people in different countries) leading to false negative matches. 

\para{Embedding-based methods} Embedding-based methods could be applied in matching. However, despite their employment on embedding cell-values, and consequently table columns \cite{fernandez2018seeping, nargesian2018table}, pre-trained models have been shown to not work well on domain-specific datasets \cite{koutras2021valentine}. On the other hand, \emph{locally-trained} embedding methods \cite{fernandez2019termite,koutras2020rema, cappuzzo2020creating} leverage the architecture of \textit{skip-gram models} \cite{mikolov2013distributed, bojanowski2017enriching} to train on corpora consisting of tabular data; yet, these still seem to be insufficiently effective when used for matching related columns \cite{koutras2021valentine}.

\para{Scalability Issues} Most importantly, applying schema matching solutions \cite{rahm2001survey,  zhang2011automatic, lehmberg2017stitching, fernandez2018seeping, koutras2020rema, cappuzzo2020creating} requires, in the worst case, computation of similarities between all pairs of columns. As the number of columns $n$ increases beyond the thousands -- a small number considering the size of data lakes and commercial databases -- computing O($n^2$) similarities is impractical.

\subsection{Dataset Discovery \& Semantics Types} 
\label{sec:rel-dd}

\para{Related-dataset search}  Related-dataset search methods \cite{fernandez2018aurum, cafarella2009data, sarma2012finding,nargesian2018table, zhu2019josie, bogatu2020dataset, zhang2020finding, fan2022semantics, khatiwada2022santos, bogatu2022voyager} rely on the syntactic-, distribution- or even embedding-similarity of data instances within dataset columns. In order to scale, related dataset search methods make use of LSH \cite{fernandez2018aurum, nargesian2018table, bogatu2020dataset} or inverted \cite{zhu2019josie} indexes. However, their application to the matching problem is not straightforward: dataset search methods return the top-k related datasets to a  dataset given as query. In the case of matching across data silos, we are not concerned with capturing the top-k related datasets, but \textit{matches among columns} across silos. Therefore, to use a related-dataset search method for capturing all possible column matches, one would need to set k to a high value, in order to expand the range of the results. Yet, this could have a very negative impact on precision, as large k values could severely increase false positive rates.

\para{Column-type detection} 
Solving the problem of matching across data silos as a column-type detection problem  \cite{hulsebos2019sherlock,zhang13sato}, assumes knowledge of the exact set of semantic types that exist across the data silos, and requires massive training data that are tailored to those types. None of these assumptions hold true in the context of data silos. Despite their proven effectiveness on column classification tasks, these methods are not applicable on silos, since the semantic types and their number are unknown when trying to find links among datasets from different silos. 

\section{Problem Definition}
\label{sec:matching}
%\asterios{We should list here what relatioships we aim at. This is the best place.}
In this work, we are interested in the problem of capturing relevance among tabular datasets that belong to different silos; we focus on tabular data since they constitute the main form of useful, structured datasets in silos and include web tables, spreadsheets, CSV files and database relations. 
% Formally, we assume a set of data silos $\mathcal{S} = \{S_1, S_2, \dots, S_n\}$, where each $S_i$ stores a set of tables $\mathcal{T}_i = \{T^1_i, T^2_i, \dots, T^{t_i}_i\}, t_i = |\mathcal{T}_i|$. In addition, each table $T^j_i$ is associated with a set of several columns $\mathcal{C}^j_i = \{C^{j,1}_i, C^{j,2}_i, \dots, C^{j, c_{ji}}_i\}, c_{ji} = |\mathcal{C}^j_i|$. Our goal is then to capture potential relationships among columns of tables that belong to different silos, i.e., return all pairs $(C^{k,l}_i, C^{m,o}_j)$, for which $i \neq j, k \neq m, l \neq o$ and $r(C^{k,l}_i ,C^{m,o}_j) = 1$. Function $r: \mathcal{C} \times \mathcal{C} \rightarrow \{0, 1\}$, where $\mathcal{C}$ is the set of all columns over all data silos, signals relatedness between a pair of columns and is known for pairs of columns coming from datasets of the same silo. 
%------------------------Original up----------------------------------%
To prepare our problem setting, we start with the following definitions. 
% \rihan{why not make all basic defs together and crispy $\rightarrow$ e.g., dataset relationships can be modeled as graphs, put RG def here and we solve the problem as graph, then start tell the narrative of GNN; do we really need so complicated defs here, even it is such a simple thing; shall we just say we are schema matching using supervised learning?}
% \vspace{-0.1cm}
%\begin{definition}[\textbf{Data silos}]
%Consider a set of \emph{data silos} $S=\{S_1, S_2, \dots, S_n\}$. 
%Each data silo $S_i\ (i \in [1, n])$ stores a set of tables.
% and  the total number of columns in these tables  is $m$.
%We denote a column from data silo $S_i$ as as $c^{i}_l$. 
% \ (l \in [1, m])$. 
% \label{def:ds}
%\vspace{-0.1cm}
%\end{definition}

\begin{definition}[\textbf{Data silos}]
Consider a set of \emph{data silos} $S=\{S_1, S_2, \dots, S_n\}$. 
Each data silo $S_i\ (i \in [1, n])$ consists of a set of tables.
Assuming that  the number of columns in $S_i$ is $d$, we denote a column from data silo $S_i$ as $c^{i}_l\ (l \in [1, d])$.
\label{def:ds}
% \vspace{-0.1cm}
\end{definition}

\begin{definition}[\textbf{Intra-relatedness and Inter-relatedness}] 
If two columns $c^{i}_k, c^{i}_l$ are from the same data silo $S_i$ ($k \neq l$), and  represent the same semantic type, we refer to their relationship as \emph{intra-related}; if two columns $c^{i}_l$ and $c^{j}_t$ ($i \neq j$) are located in different data silos, and represent the same semantic type, we refer to their relationship as \emph{inter-related}.
\label{def:rel}
% \vspace{-0.2cm}
\end{definition}

Intra- and inter-related columns refer to three different notions of matches: \emph{i}) columns that share exact value overlaps and draw values from the same domain, i.e., they are \emph{equi-joinable}, \emph{ii}) columns that share semantically equivalent values of different formats and belong to the same domain, i.e., they are \emph{fuzzily-joinable}, and \emph{iii}) columns that do not share any kind of value overlaps but store instances from the same domain,  i.e., they are \emph{unionable}.

Given a set of data silos $\mathcal{S}$, we refer to the set of all columns in $\mathcal{S}$ as $\mathcal{C}$.
For example, in Figure \ref{fig:silos} we have $\mathcal{S}=$ \{\emph{Insurance, Banking, Public District Data} \}, and the total number of columns $|C|$ is 21. 
In this work, we assume that the intra-relatedness in each data silo is known, which is common in organizations as discussed in Section~\ref{sec:intro}.

\para{The Problem of Matching Across Data Silos} Consider a set of data silos $\mathcal{S}$, and that the intra-relatedness relationships in each data silo $S_i \in \mathcal{S}$ are known. The problem of \emph{matching across data silos}, is to capture the potential inter-relatedness relationships  among the table columns belonging to different silos. 

For instance, in Figure \ref{fig:sima} we know that in the silo \texttt{Insurance} the columns \texttt{Customer.Address} and \texttt{Insurance.Address} are related. 
Now we aim to discover inter-relatedness between different silos, such as \texttt{Insurance.Address} and \texttt{District.Name} (in the silo  \texttt{Public District Data}), which are from two data silos and remain unknown among their corresponding stakeholders. In Section~\ref{sec:training} we will elaborate on how we transform the above problem to a \emph{link prediction} problem. 

\section{GNNs for Matching Data Silos}
\label{sec:cas}

%\asterios{We need to cover those reviews claiming that this is a simple applications of GNNs to an old problem. Two things here: 1) this is a new problem - I feel we did not advertise enough. The problem should, of course, pose some restrictions e.g., we cannot allow data to move from silo to silo but some profiles can. Note that the problem definition \emph{should} include a silo-type restriction.; 2) there are alternative ways of representing the problem in GNNs (even stupid ones e.g., tables are nodes, everything goes on one graph, or on separate graphs, the connected components story, etc.). We have to talk about those alternatives and show that "applying GNNs to the problem" is non trivial and that we have tried and failed.}

% All previously discussed solutions have something in common: they do not use the existing knowledge in every data silo in order to proceed. Indeed, traditional matching methods do not take into consideration the relationships that already hold among columns of different tables. In this section, we present how \method{} utilizes this knowledge by employing Graph Neural Networks (GNNs) in order to provide with inter-silo link suggestions. Towards this direction, we first introduce GNNs and describe how \method{} employs them for federating data silos.
%------------------------Original up----------------------------------%
\begin{table}[t!]
\centering
\caption{Essential notations used in the paper}
\label{tbl:nota}
\begin{tabular}{|l|l|}
\hline
\textbf{Notations} & \textbf{Description}  \\ \hline \hline
$\mathcal{G}$                     & A graph                                    \\ \hline
$v$                               & A node in $\mathcal{G}$  \\ \hline
$\mathcal{N}_v$                   & The set of neighborhood nodes of  $v$  \\ \hline
\textbf{h}$_v$                    &  Features associated with $v$   \\ \hline
$\mathbf{h}^0_v$                  & The initial feature vector of $v$ \\ \hline
$k$                               &  The layer index  \\ \hline
$\textbf{h}^k_v$                  & The $k$-th layer feature vector of $v$ \\ \hline
$\textbf{h}^{k}_{\mathcal{N}_v}$  & The $k$-th layer feature vector of $\mathcal{N}_v$   \\ \hline
\textbf{W}$^k$                    & The weight matrix to the $k$-th layer   \\ \hline \hline
$\mathcal{S}$                               &  The set of data silos \\ \hline
$i$                               &  The data silo index  \\ \hline
$S_i$                  & The $i$-th data silo \\ \hline 
$\mathcal{RG}$                  & The set of relatedness graphs of $\mathcal{S}$ \\ \hline 
$RG_i$                  & The relatedness graph of $S_i$ \\ \hline 
$\textbf{f}_v$                  & Initial feature vector of $v$ obtained via profiling\\ \hline 
$PE_i$                            & The set of positive edges of $RG_i$        \\ \hline
$NE_i$                            & The set of negative edges of $RG_i$        \\
\hline
% 
                    %   &                     \\ \hline                       
                    %   &                     \\ \hline
\end{tabular}
% \vspace{-0.3cm}
\end{table}
%The common pitfall of  solutions is the lack of utilizing existing knowledge in every data silo. 
%Indeed, the plethora of schema matching methods do not take into consideration the relationships that already hold among columns of different tables (apart from methods like \cite{li2000semint, doan2003learning} that are not applicable in our setting), which hinders the discovery of meaningful inter-related columns.

 In this section, we present how \method{} utilizes intra-silo column relatedness knowledge and manages to leverage \emph{Graph Neural Networks (GNNs)} to provide with inter-silo link suggestions. Towards this direction, we first give a preliminary introduction on  GNNs in Section~\ref{sec:gnn}.
 Then in Section~\ref{sec:model} we showcase how we model a set of data silos as graphs, and obtain the initial features via profiling in Section~\ref{sec:profile}. 
We transform the problem of matching across data silo to a \emph{link prediction} task, and describe how \method{} employs GNNs to solve the problem in Section~\ref{sec:training}. We explain \method's  algorithmic pipeline in Section~\ref{sec:pipeline}.
In Table~\ref{tbl:nota} we summarize the notations frequently used in this paper. 
% \subsection{Modeling Data Silos for GNNs}
\subsection{Preliminary: GNNs}
\label{sec:gnn}

Recently, \textit{Graph Neural Networks} \cite{wu2020comprehensive} have gained a lot of popularity due to their straightforward applicability and impressive results in traditional graph problems such as \textit{node classification} \cite{kipf2016semi, hamilton2017inductive}, \textit{graph classification} \cite{errica2019fair} and \textit{link prediction} \cite{zhang2018link, ying2018graph, fan2019graph}.  Intuitively, GNNs can learn a ``recipe" to incorporate the neighborhood information and the features of each node in order to embed it into a vector. 

In this work, we aim at finding a learning model that can perform well, not only on silos with known column relationships, but also on \textit{unseen} columns in unseen data silos. This requires a generic, inductive learning framework.
Based on the wealth of literature around GNNs, we opt for the seminal GNN model of GraphSAGE \cite{hamilton2017inductive}, which is one of the representative models 
% GNNs benefit from the features of the nodes and the relationship information among them in the graph in order to build 
generalizable to unseen data during the training process.
More specifically, GraphSAGE incorporates the features associated with each node $v$ of a graph, denoted by \textbf{h}$_v$, together with its neighborhood information $\mathcal{N}_v$, in order to learn a function that is able to embed graph nodes into a vector space of given dimensions. The embedding function is trained through message passing among the nodes of the graph, in addition to an optimization objective that depends on the use case. Typically, GraphSAGE uses several \emph{layers} for learning how to aggregate messages from each node's neighborhood, where in the $k$-th layer it proceeds as follows for a node $v$:

\begin{equation}
\begin{gathered}
\textbf{h}^{k}_{\mathcal{N}_v} = \text{AGGREGATE}_{k}(\{\textbf{h}^{k-1}_{u}, \forall u \in \mathcal{N}_v\}) 
\\ 
 \textbf{h}^{k}_{v} \leftarrow \sigma\big(\textbf{W}^{k} \cdot \text{CONCAT}(\textbf{h}^{k-1}_{v}, \textbf{h}^{k}_{\mathcal{N}_v} ) \big)
\end{gathered}\end{equation}

\vspace{2mm}

Given a node $v$, 
% to compute its $k$-th layer feature vector $\textbf{h}^k_v$, 
GraphSAGE first aggregates the representations of its neighborhood nodes from the previous layer $k$-1, and obtains $\textbf{h}^{k}_{\mathcal{N}_v}$.  
Then the concatenated (\text{CONCAT}) result of the current node representation  $\textbf{h}^{k-1}_{v}$ and the neighborhood information $\textbf{h}^{k}_{\mathcal{N}_v}$ is combined with the $k$-th layer weight matrices \textbf{W}$^k$. 
After passing the activation function $\sigma(\cdot)$, we obtain the feature vector of $v$ on the current layer $k$, i.e., $\textbf{h}^{k}_{v}$.
% and \textbf{h}$^0_v$ is node's $v$ initial feature vector.
Such a process starts from the initial feature vector of the node $v$, i.e., $\mathbf{h}^0_v$.
By stacking several such layers GraphSAGE controls the depth from which this information arrives in the graph. For instance, $k = 3$ indicates that a node $n$ will aggregate information until 3 hops away from $n$.
% \rihan{For us the above explanation is sufficient. But I worry whether someone without GNN knowledge can comprehend it easily. E.G., What is CONCAT?}

% \subsection{Our Modeling of Data Silos for GNNs}
\subsection{Modeling Data Silos as Graphs}
\label{sec:model}
We see that applying a GNN model on a given graph is seamless and quite intuitive: nodes exchange messages with their neighborhood concerning information about their features, which is then aggregated to reach their final representation. Yet, for the GNN to function properly, the graph on which it is trained should reveal information that is correct, namely we should be sure about the edges connecting different nodes.

\begin{figure}
    \centering
    \includegraphics[width=.95\columnwidth]{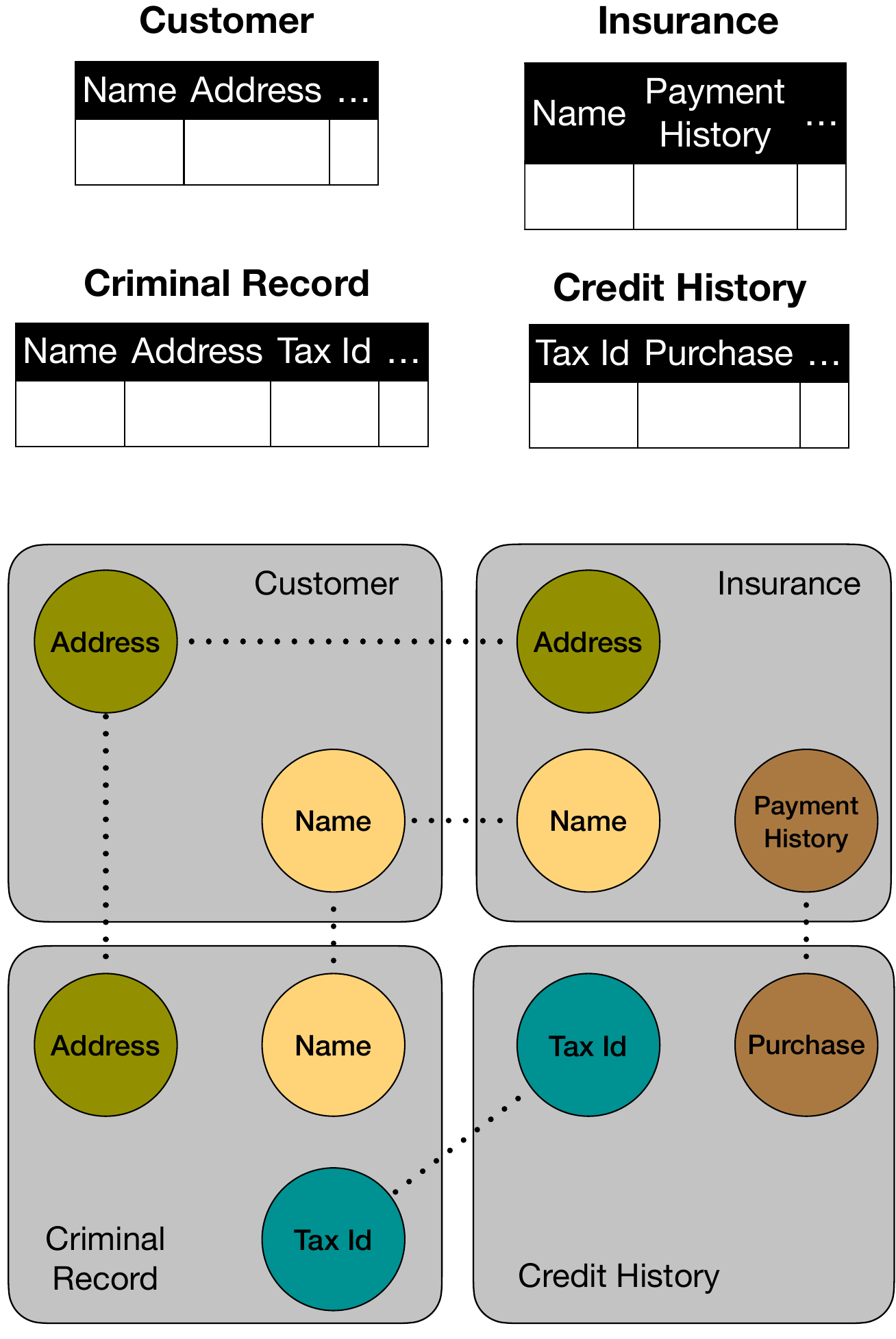}
    \caption{Relatedness graph of the \emph{Insurance} data silo.}
    \label{fig:relatedness-graph}
\end{figure}

% Based on this last observation and on the fact that data silos maintain information about relationships among their own datasets, we see that if we model each silo as a graph then this could enable the application of GNNs. In order to do so, for each data silo $S_i$, as defined in Section \ref{sec:matching}, we construct a \emph{relatedness graph} $RG_i(V_i, E_i)$, where each node $v \in V_i$ represents a column belonging to one of the tables of the silo, while each edge $e \in E_i$ denotes relatedness between the corresponding nodes (the relatedness graph is undirected). In essence, each such graph represents the links among the various tabular datasets that reside in the corresponding data silo. In Figure \ref{fig:relatedness-graph}, we see how the Insurance data silo from \autoref{fig:silos} is transformed to the corresponding relatedness graph. Based on it, we see that a silo's relatedness graph consists of several connected components, where each of them represents a different domain to which columns of the datasets that are stored in the data silo belong; thus, the neighborhood of each node in the graph includes only the nodes that are relevant to it in the silo. This is shown in Figure \ref{fig:relatedness-graph}, where we see four different connected components, colored differently, which represent four different domains in the silo: addresses, names, tax ids and purchase info. 
%------------------------Original up----------------------------------%
Based on this last observation and on the fact that data silos maintain information about relationships among their own datasets, we see that if we model each silo as a graph then this could enable the application of GNNs. In order to do so, for each data silo $S_i$, as defined in Section~\ref{sec:matching}, we construct a \emph{relatedness graph} that represents the links among the various tabular datasets that reside in the corresponding data silo.
% \rihan{Actually it is an intra-relatedness graph.}
\begin{definition}[\textbf{Relatedness graph}] Given a data silo $S_i$, its \emph{relatedness graph} $RG_i = (V_i, E_i)$ is an undirected graph with nodes $V_i$ and edges $E_i$. Each column $c^i_l$ of $S_i$ is represented as a node $v \in V_i$. For each pair of columns  $c^{i}_l, c^{i}_t$ of $S_i$ that are intra-related, there is an edge $e \in E_i$ between their corresponding nodes in $RG_i$.   
\label{def:rg}
\vspace{-0.1cm}
\end{definition}
For example,  Figure \ref{fig:relatedness-graph}  shows the corresponding relatedness graph of the  data silo \emph{Insurance} from \autoref{fig:silos}. 
Based on it, we see that a silo's relatedness graph consists of several connected components, where each of them represents a different domain to which columns of the datasets that are stored in the data silo belong; thus, the neighborhood of each node in the graph includes only the nodes that are relevant to it in the silo. This is shown in Figure \ref{fig:relatedness-graph}, where we see four different connected components, colored differently, which represent four different domains in the silo: addresses, names, tax ids and purchase info. 

\subsection{Profiles as Initial Features} 
\label{sec:profile}
\para{Initialization requirement of GNNs} With \method{} we opt for applying the GraphSAGE model using the relatedness graphs of the corresponding data silos. For this to be possible two conditions should be satisfied about the relatedness graph: \emph{i}) there should be a representative set of edges and \emph{ii}) each node should come with an initial feature vector. \method's relatedness graphs already satisfy the first condition, since every such graph includes edges denoting similar columns. Yet, nodes in the relatedness graphs are featureless. Moreover, in order to leverage  GNNs and use them for matching across data silos, we need to employ them towards a specific goal. Therefore, in the following we discuss how to produce initial features for each column-node in a relatedness graph, and present a method of using a GNN model for bridging data silos by modeling our problem as a link prediction task.

% \para{Profiles as initial features} 
\para{Initial feature vectors from data profiles}
 In order to handle the feature initialization requirement, in \method{} we draw inspiration from the \emph{data profiling} literature \cite{abedjan2015profiling}. In the case of tabular data, profiles summarize the information of a data element, by calculating a series of simple statistics (e.g. number of null values, aggregates etc.). Consequently, we can utilize a simple profiler in order to associate each column in a data silo to a feature vector, summarizing statistical information about it. 

In specific, for each data silo, we feed all the including tables into the profiling component we adopt from \cite{hulsebos2019sherlock}. However, since we need the initial profiles to summarize simple information for each column (so as not to depend on complex profiles), we exclude the features referring to pre-trained value and paragraph embeddings. In short, \method{} computes a feature vector for each column in a silo by collecting the following:%\footnote{Due to space restriction, we will release the full list of applied features and the results online.}:

\parait{-- Global statistics} Those include aggregates on high level characteristics of a column, e.g. number of numerical values among the included instances.

\parait{-- Character-level distributions} For each of the 96 ASCII characters that might be present in the corresponding values of the column, we save charachter-level distributions. Specifically, the profiler counts the number of each such ASCII character in a column and then feeds it to aggregate functions, such as \emph{mean}, \emph{median} etc.

Using the above profiling scheme, we associate each node $v$, belonging to a relatedness graph $RG_i$, with a vector $\textbf{f}_v$. This $\textbf{f}_v$ will serve as $\textbf{h}^0_v$ for initializing the feature vector of $v$ before starting the GraphSAGE training process, as shown in \autoref{fig:sima}b.

\section{Training GNNs for Matching Silos}
\label{sec:training}
\para{Matching across silos as link prediction} In order to leverage the capabilities of a GNN, there should be an objective function tailored to the goal of the problem that needs to be solved. With \method{} we want to be able to capture relatedness for every pair of columns belonging to different data silos, which translates to the following objective.
\begin{problem} [\textbf{Link prediction of relatedness graph}]
	\label{def:gnn_lp} 
Consider a set of relatedness graphs $\mathcal{RG}$, the challenge of \emph{link prediction across relatedness graph}s is to build a model $\mathcal{M}$ that \emph{predicts} whether there should be an edge between nodes from different relatedness graphs. 
Given a pair of nodes $(u,v)$ from two different relatedness graphs $RG_i, RG_j \in \mathcal{RG}$ ($i \neq j$) where $u \in RG_i$, $v \in RG_j$,  ideally 
\begin{align*}
 		\begin{split} 
 		\mathcal{M}(u , v) =  \begin{cases}1, &u\  and\  v\ are\  linked  \\0, & otherwise\end{cases}
 		\end{split}
\end{align*}
\end{problem}

It is easy to see that we have now transformed our initial matching across data silos problem to a  \emph{link prediction} problem over the relatedness graphs.
% Towards this direction, we train a prediction function $\phi$ that receives as input the representations $\textbf{h}_u$ and $\textbf{h}_v$, of the corresponding nodes $u$ and $v$, from the last layer of the GraphSAGE neural network, and computes a similarity score $sim(u,v) = \phi(\textbf{h}_u, \textbf{h}_v)$. Nonetheless, the training samples we get from our relatedness graphs contain only pairs of nodes for which a link should exist (\emph{positive edges}). Thus, we need to provide the training process with a corresponding set of \emph{negative edges}, which connect nodes-columns that are not related. In order to do so for every relatedness graph $RG_i$ we construct a set of negative edges $NE_i = \{(u, v) | r(u, v) = 0 \wedge u,v \in RG_i\}$, since we know that nodes belonging to different connected components in $RG_i$ represent pairs of unrelated columns in the corresponding data silo $S_i$; we elaborate on such negative edge sampling strategies in Section \ref{sec:sampling}. After constructing the set of negative edges, we initiate the training process with the goal of optimizing the following the \emph{cross-entropy} loss function:
%------------------------Original up----------------------------------%

\para{Two types of edges for training} Towards this direction, we train a prediction function $\phi$ that receives as input the representations $\textbf{h}_u$ and $\textbf{h}_v$, of the corresponding nodes $u$ and $v$, from the last layer of the GraphSAGE neural network, and computes a similarity score $sim(u,v) = \phi(\textbf{h}_u, \textbf{h}_v)$. 

To train a robust GNN model, we need the following two types of edges in our relatedness graph.
\begin{definition}[\textbf{Positive edges and negative edges}]
In a relatedness graph $RG_i = (V_i, E_i)$, we refer to each edge $e \in E_i$ as a \emph{positive edge}; if a `virtual' edge $e$ connects two unrelated nodes $u$ and $v$, we refer to it as a \emph{negative edge}. 
Thus, we obtain the following two sets of edges.
\begin{center}
 Positive edges $PE_i = \{(u, v) | r(u, v) = 1 \wedge u,v \in RG_i\}$ 
 
 Negative edges $NE_i = \{(u, v) | r(u, v) = 0 \wedge u,v \in RG_i\}$ 
\end{center} 
\label{def:pos_neg_edge}
\vspace{-0.1cm}
\end{definition}

To differentiate with negative edges $NE_i$, in the sequel we refer to the edges of a relatedness graph $V_i$ as positive edges $PE_i$.
Notably, the training samples we get from our relatedness graphs contain only pairs of nodes for which a link should exist (i.e., positive edges $PE_i$). Thus, we need to provide the training process with a corresponding set of negative edges, which connect nodes-columns that are not related. To do so, for every relatedness graph $RG_i$ we construct a set of negative edges $NE_i$, since we know that nodes belonging to different connected components in $RG_i$ represent pairs of unrelated columns in the corresponding data silo $S_i$; we elaborate on negative edge sampling strategies in Section~\ref{sec:sampling}.

\para{Two-fold GNN model training} After constructing the set of negative edges, we initiate the training process with the goal of optimizing  the following \emph{cross-entropy} loss function:

\begin{align}
\begin{split}
    \mathcal{L} &= - \sum_{(u,v) \in RG_i}{\log{\sigma(sim(u,v))}} \\
   &\quad - \sum_{(u,v) \in NE_i}{\big[1-\log(\sigma(sim(u, v)))\big]}
\end{split}
\end{align}

\noindent where $\sigma(\cdot)$ is the sigmoid function and $1 \leq i \leq n$, with $n$ representing the number of relatedness graphs (constructed from the original data silos) included in training data. The similarity scores are computed by feeding pairs of node representations to a \emph{Multi-layer Perceptron} (MLP), whose parameters are also learned during the training process in order to give correct predictions. Intuitively, with this model training we want to compute representations, so as to build a similarity function (through the training of the MLP), which  based on them, correctly distinguishes semantically related from unrelated nodes-columns.
%\asterios{in the above, I would like to see more intuition about what we are doing. We are presenting things in a very "dry" fashion, and although I do understand what is said, I am missing my compass: the why/big picture.}

\vspace{1mm}
\noindent To summarize, \method{} uses a two-fold model, which consists of:

\begin{itemize}
    \item A GraphSAGE neural network that applies message passing and aggregation (Equation 1) in order to embed the nodes-columns of the relatedness graph into a vector space of given dimensions.
    \item A MLP, with one hidden layer, which receives in its input a pair of node representations and based on them it calculates a similarity score in order to predict whether there should be a link or not between them, i.e., whether the corresponding columns are related.
\end{itemize}

\RestyleAlgo{ruled}
\begin{algorithm}[tp]
\SetAlgoLined
\caption{SiMa}
\label{alg:casanova}

\SetKwInOut{Input}{Input}
\SetKwInOut{Output}{Output}
\SetKwComment{Comment}{// }{}
\SetKw{KwBy}{by}
\SetKwFunction{FindCompress}{FindCompress}

\Input{Set of data silos $\mathcal{S}$ \\ Model $\mathcal{M}$\\ Profiler $\mathcal{P}$\\Number of training epochs $e$ \\ }
\Output{Trained model $\mathcal{M}$}

$\mathcal{RG} \leftarrow \{\}$ \tcp{Initialize set of relatedness graphs}
$\text{f} \leftarrow []$ \tcp{List of initial node feature vectors}

$n \leftarrow |\mathcal{S}|$

\For{$i\gets1$ \KwTo $n$}{
    $RG_i \leftarrow \text{ConstructGraphFromSilo}(S_i)$ 
    
    $\mathcal{RG}.\text{add}(RG_i)$ 
    
     \ForEach{node $u\in RG_i$}{
        \textbf{f}$_u$ $\leftarrow \mathcal{P}(u)$ \tcp{Compute profile of corresponding column and store it as u's initial feature vector}
        f.append(\textbf{f}$_u$) 
     }
  }

$\mathcal{PE}, \mathcal{NE} \leftarrow \{\}$ \tcp{Initialize sets of positive/negative edges}

\For{$i\gets1$ \KwTo $n$}{
    \ForEach{edge $(u, v)\in RG_i$}{$\mathcal{PE}.\text{add}((u, v))$}
    
    $\mathcal{NE}.\text{union}(\text{SampleNegativeEdges}(RG_i))$  
   }

\For{$i\gets1$ \KwTo $e$}{
    \textbf{h} $\leftarrow \mathcal{M}$.GraphSage($\mathcal{RG}$, f) \tcp{Apply GraphSAGE to all relatedness graphs and get node embeddings}
    $PosEdgePred \leftarrow \mathcal{M}.$MLP($\mathcal{PE}$, \textbf{h})  \tcp{Get link predictions for positive edges}
    
    $NegEdgePred \leftarrow \mathcal{M}.$MLP($\mathcal{NE}$, \textbf{h}) \tcp{Get link predictions for negative edges}
    
    Loss$\leftarrow$ComputeLoss($PosEdgePred$, $NegEdgePred$) \tcp{Compute cross-entropy loss based on predictions}
    
    Loss.BackPropagate($\mathcal{M}$.parameters) \tcp{Tune model parameters with backwards propagation}
    }
\end{algorithm}

In the above model, there can be certain modifications with respect to the kind of GNN used (e.g. replace GraphSAGE with the classical Graph Convolutional Network \cite{kipf2016semi}) and prediction model (e.g. replace MLP by a simple dot product model). However, since the focus of this work is on building a method which uses GNNs as a tool towards matching across data silos, and not on comparing/proposing novel GNN-based link prediction models, we opt for a model architecture similar to the ones employed for link prediction \cite{vretinaris2021medical, ying2018graph}. 

\subsection{\method's Pipeline}
\label{sec:pipeline}

% In Algorithm \ref{alg:casanova} we show the pipeline that we employ with \method, towards building a model that is able to represent columns of data silos in such a way, so that relatedness prediction based on them is correct.
In Algorithm \ref{alg:casanova} we show the pipeline that we employ with \method. 
The key challenge here is to build a model that can represent columns of data silos in such a way, so that relatedness prediction based on them is correct.
Our method has four inputs: \emph{i}) the set of data silos $\mathcal{S}$, \emph{ii}) our defined model $\mathcal{M}$, including the GraphSAGE neural network and the MLP predictor,  \emph{iii}) the profiler $\mathcal{P}$ that we use in order to initialize feature vectors of nodes, and \emph{iv}) the number of training epochs $e$. The output of \method{} consists of the trained model $\mathcal{M}$, which can then be used to embed any column of a data silo and, based on these embeddings, predict links between columns.

Initially, all data silos in $\mathcal{S}$ are transformed to their relatedness graph counterpart. In addition, we compute the corresponding profiles of each node and store them as initial feature vectors (lines 4-11). Based on these graphs, we construct the sets of positive and negative edges to feed our training process (lines 13-18). While getting positive edges is trivial, since we just fetch the edges that are present in the relatedness graphs, constructing a set of negative edges requires a sampling strategy (line 17). This is because the set of \textit{all} negative edges is orders of magnitude larger than the set of positive ones. Ergo, we need to sample some of these negative edges in order to balance the ratio of positive to negative examples for our training. We elaborate on our optimized strategies for negative edge sampling in Section~\ref{sec:sampling}. 

Following the preparation of positive and negative edge training samples, we move to the training of our model (lines 19 - 25). In specific, we start by applying the current GraphSAGE neural network %,so as to receive the vector representations (i.e., \textbf{h}) of all nodes included in the relatedness graphs, 
through the message passing and aggregation functions shown in Equation 1. At the next step, we get the predictions for the pairs of nodes in the set of positive and negative edges respectively (lines 21-22), by placing in the input of our defined MLP architecture their corresponding embeddings. Finally, the cross-entropy loss is calculated (Equation 2) based on all predictions made for both positive and negative edges (line 23) and based on it we back propagate the errors in order to tune the parameters of the GraphSAGE and MLP models used (line 24). The training process repeats for the number of epochs $e$, which is specified in the input. In the end of this loop, we get our trained model $\mathcal{M}$ which is able to embed columns in data silos and, based on these representations, predict whether they are related or not. %We further propose an optimized training method in Section~\ref{sec:incre_train}. 

\begin{figure*}[t!]
    \centering
    \includegraphics[width=\textwidth]{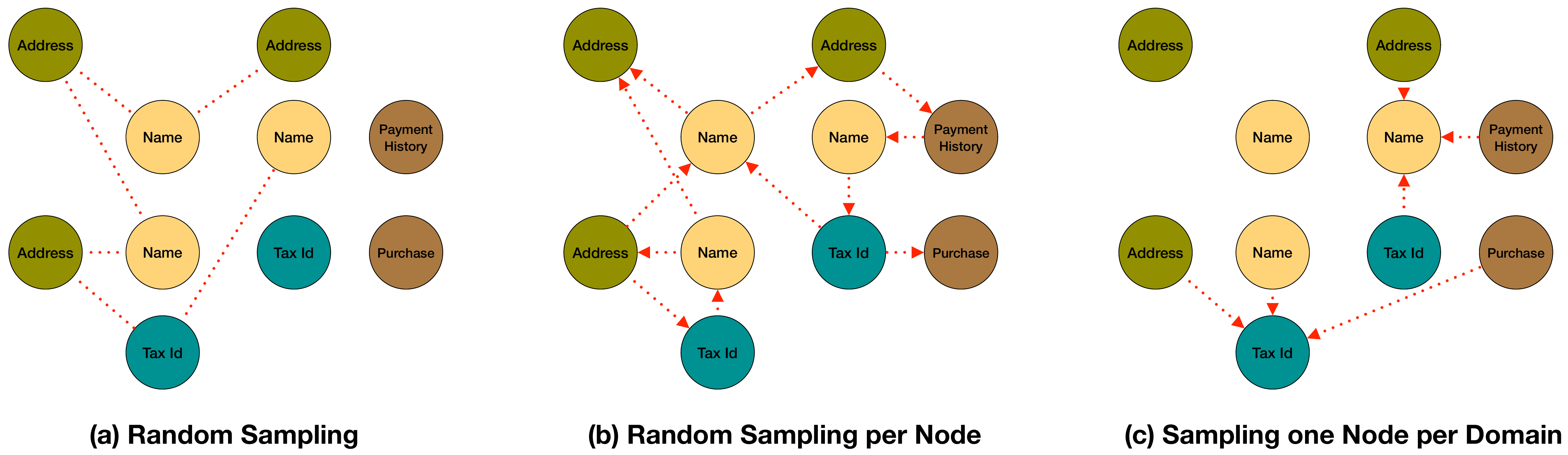}
    \caption{Strategies for negative edge sampling on the relatedness graph of the insurance data silo.}
    \label{fig:sampling}
\end{figure*}

\section{Optimization Techniques}
\label{sec:opt}

% In Section \ref{sec:pipeline}, we discussed about the pipeline we utilize in \method{} in order to build a model that predicts links among data silos. From the whole pipeline, as shown in Algorithm \ref{alg:casanova}, there are two processes that stand out and are not straightforward: \emph{i}) the negative edge sampling and \emph{ii}) model training. Thus, in this section we present three negative edge sampling techniques, each improving upon the shortcomings of the previous one. Then, we justify why training our model upon every data silo at once might not be beneficial for the effectiveness of our prediction model, and describe an incremental training scheme.
%------------------------Original up----------------------------------%
%RH: we don't so much details here, just to say they are novel+optimization
%Details about each tech will be explained in the beginning of each subsection
In this section, we present novel techniques applied in Algorithm \ref{alg:casanova}: \emph{i}) sampling (Section~\ref{sec:sampling}) and \emph{ii}) incremental model training (Section~\ref{sec:incre_train}).

\subsection{Negative Sampling Strategies}
\label{sec:sampling}

Since the number of possible negative edges in our relatedness graphs might be overwhelming with respect to the number of positive edges, we need to devise negative sampling strategies. In fact, negative sampling for \emph{graph representation learning} has been shown to drastically impact the effectiveness of a model \cite{yang2020understanding}. %However, since we do not want to add more complexity to our model, with \method{} we want to introduce sophisticated, while also effective, 
Such sampling techniques can provide with negative edge samples that help our link prediction model distinguish related/dissimilar columns. Thus, in the following we describe three negative sampling strategies (termed as NS1, NS2, NS3), each enhanced with different insights. It is important to mention here that these negative sampling techniques take place inside every relatedness graph, where we have the knowledge of which node pairs represent negative examples. In Figure \ref{fig:sampling} we depict each sampling strategy and how it operates on the relatedness graph of Figure \ref{fig:relatedness-graph}.

\para{NS1: Sampling on whole graph} The most straightforward and simple way to compute a sample of negative (non-directed) edges per relatedness graph, is to randomly sample some of them out of the set of all possible negative node pairs. Specifically, based on the node connectivity information we have about each relatedness graph, we are able to compute the full set of distinct pairs which include nodes from different connected components. Then, we randomly pick  some of them in order to construct a set with a size equal to the number of positive edges in the corresponding relatedness graph to feed to our loss function.%This way we guarantee that there is a balance between the positive and negative edges included in the training data, so as the model does not overfit. 

% In Figure \ref{fig:sampling}a we show how this sampling strategy could proceed given the relatedness graph of Figure \ref{fig:relatedness-graph}. We see that since we have six positive edges in the relatedness graph, we also randomly select six negative edges as a sample. 
% Yet, this sampling strategy has a major shortcoming: there could be nodes that are not connected with any negative edge in the sample, like the three rightmost nodes in the figure. This could severely affect the training process, since for these nodes we miss information about nodes they should not relate to. Moreover, it might be that certain nodes show up more frequently in the negative samples than others, which creates an unwanted imbalance in the training data for negative examples.
%------------------------Original up----------------------------------%
As we see in Figure \ref{fig:sampling}a, a major drawback is that there could be nodes not connected with any negative edge in the sample, like the three rightmost nodes in the figure. This could severely affect the training process, since for these nodes we miss information about nodes they should not relate to. Moreover, it might be that certain nodes show up more frequently in the negative samples than others, which creates an unwanted imbalance in the training data for negative examples. 

\para{NS2: Sampling per node} To guarantee that every node is associated with at least one negative edge, we randomly sample negative edges for each node separately. To balance the number of positive and negative edges that a node is associated with, we specify the sample size to be equal to the degree of the node in the relatedness graph, i.e., to the number of positive edges; since we want to control the number of incoming negative edges per node, we opt for directed edges. %Therefore, we make sure that the training process receives balanced information for every node in the graph.

Figure \ref{fig:sampling}b shows a possible output of such a negative edge sampling strategy. In contrast to the previous strategy, we see that now every node in the graph receives a sample of directed negative edges, of size equal to its corresponding degree in the original relatedness graph. Nonetheless, this improved sampling strategy does not ensure that a node will receive negative edges from a set of nodes that belong to different connected components, namely different column domains. For example, in Figure \ref{fig:sampling}b the upper left ``Address'' node receives two edges both coming from the connected component representing the domain of customer names. This non-diversity of the negative samples that are associated with each node, disrupts the learning process since the model does not receive enough information about which columns should not be regarded as related.

\para{NS3: Sampling per domain} To improve the shortness of diversity in the negative edges each node receives, we impose sampling per node to take place per different domain, i.e., for each different connected component in the relatedness graph. In detail, this time we pick the random samples based on each connected component that has not yet been associated to the node. Hence, each node receives at least one negative edge from every other connected component in the graph, ensuring this way that there is diverse and complete information with respect to domains that the corresponding column does not relate. To keep the number of negative edges close to the number of positive ones, we specify one random sample from each domain per node. 

To illustrate how the above strategy proceeds, in Figure \ref{fig:sampling}c we show negative samples computed only for two different nodes, ``Tax id" and ``Name" (we do so in order to not overload the figure with negative edges for all nodes). Indeed, as we discussed above, both of these nodes receive exactly one randomly picked edge from each unrelated domain. Therefore, every node has complete information which can be leveraged by our proposed model in order to learn correctly which pairs of nodes should not be linked.

\para{Remarks} When using NS3 the number of negative edges sampled for training might be considerably higher than the one of positive edges. To deal with this imbalance of positive and negative data, we use the weighted version of the binary cross-entropy function:
\vspace{-2mm}
\begin{align}
\begin{split}
    \mathcal{L} &= - \sum_{(u,v) \in RG_i}{\bm{w_p}\cdot\log{\sigma(sim(u,v))}} \\ 
    &\quad - \sum_{(u,v) \in NE_i}{\log(1 - \sigma(sim(u, v)))}
\end{split}
\end{align}

% \vspace{1mm}

\noindent where $w_p$ is the weight we use to balance the contribution of the positive and the negative examples, which we set to be equal to the ratio of negative to positive edges included in the training.

% \begin{algorithm}
% \SetAlgolined
% \caption{Incremental Training}
% \label{alg:training}

% \SetKwInOut{Input}{Input}
% \SetKwInOut{Output}{Output}
% \SetKwComment{Comment}{// }{}
% \SetKw{KwBy}{by}
% \SetKwFunction{FindCompress}{FindCompress}

% $n \leftarrow |\mathcal{RG}|$ \;
% $TG \leftarrow []$ \Comment*[r]{TG holds list of relatedness graphs included in the training}

% \For{$i\gets1$ \KwTo $n$}{
% $TG$.append($RG_i$) \;
% \For{$j\gets1$ \KwTo $e_p$}{ 
%     \textbf{h} $\leftarrow \mathcal{M}$.GraphSage($TG_i$, f$_i$) \Comment*[r]{Apply GraphSAGE only on relatedness graphs in TG}
%     $PosEdgePred \leftarrow \mathcal{M}.$MLP($\mathcal{PE}_i$, \textbf{h})  \Comment*[r]{Get link predictions for positive edges}
    
%     $NegEdgePred \leftarrow \mathcal{M}.$MLP($\mathcal{NE}_i$, \textbf{h}) \Comment*[r]{Get link predictions for negative edges}
    
%     Loss $\leftarrow$ ComputeLoss($PosEdgePred$, $NegEdgePred$) \Comment*[r]{Compute cross-entropy loss based on the predictions}
    
%     Loss.BackPropagate($\mathcal{M}$.parameters) \Comment*[r]{Tune model parameters with backwards propagation}
%     }
 
% }
% \end{algorithm}
\begin{algorithm}[t]
\SetAlgoLined
\caption{Incremental Training}
\label{alg:training}

\SetKwInOut{Input}{Input}
\SetKwInOut{Output}{Output}
\SetKwComment{Comment}{// }{}
\SetKw{KwBy}{by}
\SetKwFunction{FindCompress}{FindCompress}

$n \leftarrow |\mathcal{RG}|$ \;
$TG \leftarrow []$ \tcp{List of relatedness graphs included in the training}

\For{$i\gets1$ \KwTo $n$}{
$TG$.append($RG_i$) \;
\For{$j\gets1$ \KwTo $e_p$}{ 
    \textbf{h} $\leftarrow \mathcal{M}$.GraphSage($TG_i$, f$_i$) \tcp{Apply GraphSAGE only on relatedness graphs in TG}
    $PosEdgePred \leftarrow \mathcal{M}.$MLP($\mathcal{PE}_i$, \textbf{h})  \tcp{Get link predictions for positive edges}
    
    $NegEdgePred \leftarrow \mathcal{M}.$MLP($\mathcal{NE}_i$, \textbf{h}) \tcp{Get link predictions for negative edges}
    
    Loss $\leftarrow$ ComputeLoss($PosEdgePred$, $NegEdgePred$) \tcp{Compute cross-entropy loss based on predictions}
    
    Loss.BackPropagate($\mathcal{M}$.parameters) \tcp{Tune model parameters with backwards propagation}
    }
 
}
\end{algorithm}

\subsection{Incremental Training}
\label{sec:incre_train}
Originally, \method{} trains on the positive and negative samples it receives by taking into consideration every relatedness graph in the input (lines 19-25 in Algorithm \ref{alg:casanova}). However, proceeding with training on the whole set of graphs might harm the effectiveness of the learning process, since the model in each epoch trains on the same set of positive and negative samples; hence, it can potentially overfit. Therefore, we need to devise an alternative training strategy, which feeds the model with new training data periodically.

Towards this direction, we design an incremental training scheme that proceeds per relatedness graph. In specific, we initiate training with one relatedness graph and the corresponding positive/negative samples we get from it. After a specific number of epochs, we add the training samples from another relatedness graph and we continue the process by adding every other relatedness graph. In this way, we help the model to deal periodically with novel samples potentially representing previously unseen domains that the new relatedness graph brings; thus, we increase the chances of boosting the effectiveness that the resulting link prediction will have. Essentially, our incremental training scheme resembles \textit{curriculum learning} \cite{bengio2009curriculum} in that it constantly provides the learning process with new data; yet, curriculum learning methods also verify that the training examples are of increasing difficulty.

% \rihan{Due to the large overlap with Alg.1, it is questionable whether we should have Alg.2 separately}
Algorithm \ref{alg:training} shows how incremental training proceeds and replaces the original training scheme in the context of our initial pipeline in Algorithm \ref{alg:casanova}.  We see that the only difference with the previous scheme is that now we train the model on an incrementally growing set of relatedness graphs ($TG$ of lines 2), which is initialized with the first relatedness graph and it receives an extra graph, periodically every $e_p$ epochs (which we assume to be a fraction of the original number of epochs $e$ in Algorithm \ref{alg:casanova}), until it contains all of them in the final iteration. For each epoch, we apply GraphSAGE only on the relatedness graphs in $TG$ (lines 6) and use positive/negative samples coming from them in order to train the MLP (lines 8-9).

\section{Experimental Evaluation}
\label{sec:experiments}

\begin{figure*}[t!]
    \centering
    
    \begin{minipage}{\columnwidth}
       
            \includegraphics[width=\columnwidth]{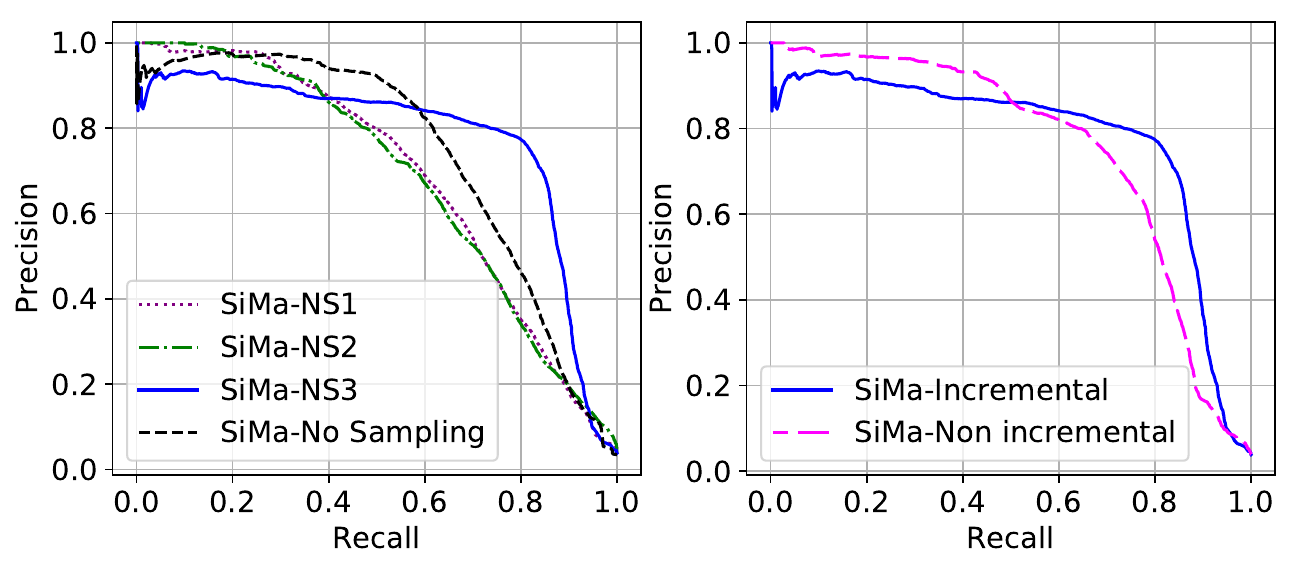}
            \vspace{-7mm}
            \subcaption{NYC OpenData}
            \label{fig:ablation_ny}
    \end{minipage}
    \begin{minipage}{\columnwidth}
             \includegraphics[width=\columnwidth]{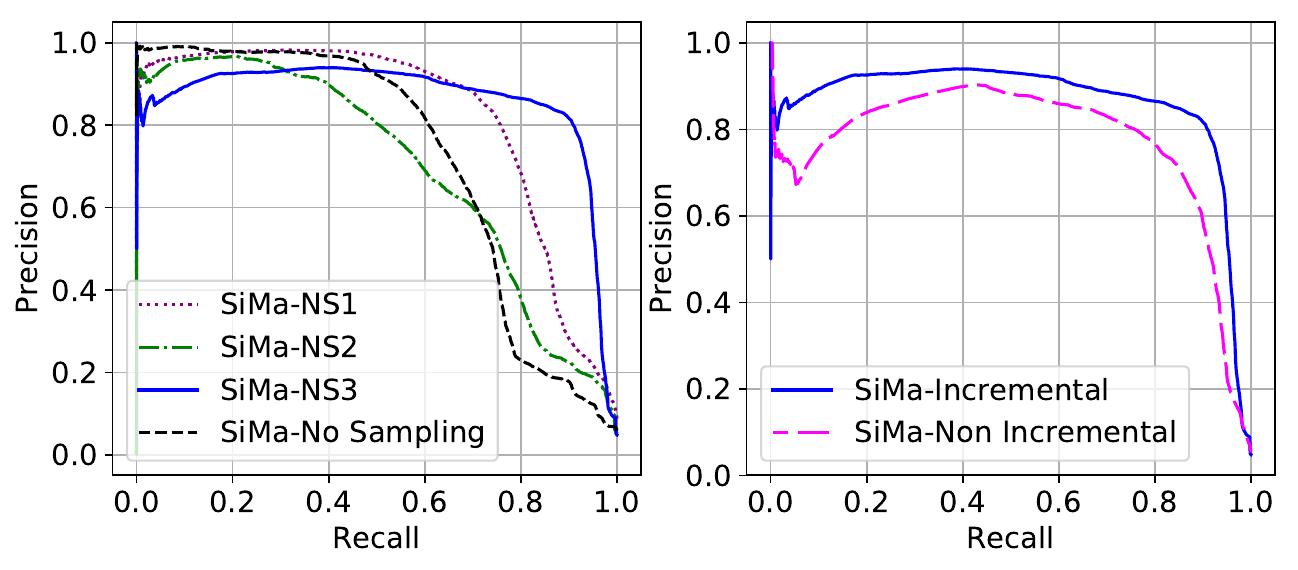}
             \vspace{-7mm}
             \subcaption{LA OpenData}
             \label{fig:ablation_la}
    \end{minipage}
    \vspace{-3mm}  
    \caption{Effect of negative edge sampling techniques and training schemes.}
    \label{fig:ablation}
    \vspace{-3mm}
\end{figure*}

In this section, we assess the effectiveness and efficiency of \method{} through an extensive set of experiments. In what follows, we first describe our experimental setup, namely the datasets, baselines and settings against which we assess our method. We then present our experimental results, where we focus on: \emph{i}) the effect of different parameters for training the GraphSAGE model, \emph{ii}) how the different sampling and training techniques (Section~\ref{sec:opt}) affect \method's effectiveness and execution time, and \emph{iii}) how \method{} compares with other matching and simple ML baselines both in terms of effectiveness and efficiency. Our main results can be summarized as follows:

\begin{itemize}
    \item The optimization techniques introduced in \autoref{sec:opt} considerably boost \method's effectiveness \autoref{fig:ablation}.
    \item \method's GNN-based model leverages existing matches better than a simple ML baseline \autoref{fig:comparison} that takes into account only the profiles.
    \item \method{} is more effective than the state-of-the-art schema matching method, due to its ability to maintain higher precision values when recall increases \autoref{fig:comparison}.
    \item Contextualized representations are not suitable for matching columns across data silos.
    \item \method{} exhibits lower execution times than other methods \autoref{tab:efficiency}. Especially when compared to state-of-the-art matching methods the gap is considerably high.
\end{itemize}

\subsection{Setup}

\para{Methods used for evaluation} We make use of three different methods for comparing \method{} in terms of effectiveness and efficiency:

\paranofull{-- COMA} \cite{do2002coma}, which is a seminal and state-of-the-art matching method that combines multiple criteria in order to output a set of possible matches. We make use of the COMA version that uses both schema and instance-based information about the datasets in order to proceed. Note that  \emph{COMA is not an applicable solution to the problem} of matching across data silos as studied in this paper (Section \ref{sec:rel-sm}). However, we use it in order to see how close \method{} can get to a state of the art matching method, on the same data and in a non-siloed setting. In our experiments we use COMA 3.0 Community Edition.

\paranofull{-- Starmie} \cite{fan2022semantics} is a state-of-the-art \emph{top-k unionable table search} in \emph{data lakes}. The method employs a multi-column table encoder that serializes instances from tables to feed them into a pre-trained \emph{Language Model} (LM) (specifically, the authors use RoBERTa \cite{liu2019roberta}). Starmie uses contrastive learning \cite{chen2020simple} to produce column representations that capture relatedness. In our evaluation, we use Starmie, as shared in a public repository\footnote{\url{https://github.com/megagonlabs/starmie}}, to produce contextualized column representations for the columns of the datasets included in each silo. We then compute pairwise cosine similarity for columns among datasets of different silos. We ran Starmie with default parameters, except for tuning the max sequence length to 256, number of epochs to 5, and batch size to 8.

\paranofull{-- Baseline: MLP}. To show the gains of using \method's GNN model to represent dataset columns, we compare our method against a simple Multi-Layer Perceptron (MLP) prediction model. In specific, we use a MLP with one hidden layer, which receives in its input pairs of profiles and learns to predict whether there is a relationship among the columns they represent. In other words, this model disregards column representations computed through \method's GNN model and straightforwardly uses only column profiles and information about existing column matches.

 Note that we have not included baselines from the dataset discovery literature \cite{fernandez2018aurum, cafarella2009data, sarma2012finding,nargesian2018table, zhu2019josie, bogatu2020dataset, zhang2020finding} as those do not address the problem of matching columns across silos; instead, they address the top-k similar dataset retrieval problem, which makes them not directly applicable to our problem setting.

Except for COMA, Starmie and the MLP baseline, we ran our experiments with two additional schema matching techniques, namely \emph{Distribution-based matching} \cite{zhang2011automatic} and \emph{EmbDI} \cite{cappuzzo2020creating}. Both the Distribution-based method and EmbDI exhibited consistently worse results than COMA, exhibiting very high false negative rates. In addition, EmbDI can only run on considerably smaller number of datasets than the ones we examine in our evaluation. Therefore, we omit presentation of results coming from these methods, and keep COMA as the representative state-of-the-art  matching method.
  
%Except for COMA, DB and the MLP-Matcher, we ran our experiments with two additional techniques, namely \emph{EmbDI} \cite{cappuzzo2020creating} and \emph{Starmie} \cite{fan2022semantics}; yet, both exhibited consistently low precision even for low recall values. In fact, EmbDI can only run on considerably smaller number of datasets than the ones we examine in our evaluation. On the other hand, the contextualized column representations (through BERT \cite{devlin2018bert}) we got through Starmie seemed to output a considerably high number of false negatives, when used to predict matches through cosine similarity. Therefore, we omit presentation of results coming from these methods, and keep only COMA and DB as the representative state-of-the-art methods. %We discuss these results in a comprehensive technical report \cite{techreport}.

\para{Real-world Datasets and Ground Truth} Since there is no benchmark available for matching across silos in the area of schema matching, nor in the area of related dataset search, we opted for leveraging two real-world, open data repositories:

\begin{itemize}
    \item \textbf{NYC OpenData}. The New York City OpenData repository\footnote{\url{https://opendata.cityofnewyork.us/}} contains public data published by New York City agencies and other partners. For the needs of this paper, we use tables under the \emph{City Government} category.
    \item \textbf{LA OpenData}. The Los Angeles OpenData portal\footnote{https://data.lacity.org/} encompasses public datasets covering different activities and sectors of the city of Los Angeles. We focus on tabular data under the \emph{Administration \& Finance} category.
\end{itemize}

For each of the above sources we select a subset of tables and curate them to contain only columns that \emph{i}) store categorical or text data, and \emph{ii}) the majority of their instances are not null values. Matching columns that store numerical data is out of the scope of this paper: calculating distribution similarity or set overlaps would be the adequate method to use in such cases \cite{zhu2019josie}. Based on this curation, we ended up with 22 tables, for which we manually annotated column matches among them, including equi/fuzzy-joins as well as columns of the same domain that are non-overlapping. We captured 125 and 193 column matches for the tables from the NYC and LA OpenData repositories, respectively.

To derive a larger number of tables, we adopt the method of \cite{nargesian2018table, lee2007etuner}. These methods produce pairs of tabular datasets that share a varying number of columns/rows. We make use of Valentine~\cite{koutras2021valentine} to create scenarios of equi-joinable, fuzzily-joinable and unionable columns of varying difficulty (i.e., zero/low/high - exact/non-exact value overlaps), based on the subsets of tables we created from the two OpenData repositories.

\para{Creating Data Silos} To evaluate our method we construct sets of data silos based on the tables we derived from the two OpenData repositories. Particularly, we create two benchmarks containing a given number of data silos, where each silo contains a number of tables coming from different source tables. This way we ensure that there is a sufficient amount of column matches within each silo from which our model can learn to predict column relationships among datasets across different silos. In \autoref{tab:silo_config} we detail the number of tables, column matches inside (used for training) and among (used for testing) silos for each of the two benchmarks we created.

\para{Effectiveness calculation} We evaluate the effectiveness of \method{} and the methods we compare against, by computing precision-recall curves based on the predictions (similarity scores for the case of COMA) that we retrieve for every possible pair of columns belonging to datasets of different silos. We opt for using \textit{precision-recall curves}. Those are ideal for showing effectiveness results, when the distribution of labels in the test set is considerably imbalanced \cite{he2009learning}, which is the case in our benchmarks. Indeed, in realistic matching scenarios, the number of non-matching column pairs, significantly outnumbers the matching column pairs. Notably, with precision-recall curves we show the effectiveness of methods with respect to varying similarity thresholds. Therefore, we achieve a non-biased presentation of results across the board, in contrast to showing precision and recall values only for specific threshold values.

\para{Implementation details} We experimented with different parameters for training \method's model and the simple MLP-baseline to pick the configuration that performs the best in both benchmarks. We list our observations in the following:

\paranofull{-- GNN layers:} \method{} does not benefit from using more than one layers for GraphSAGE, since the connected components formed by our graph construction method are complete graphs. Moreover, we found out that using \textit{max-pooling}, as described in \cite{hamilton2017inductive}, to aggregate the representations of each node's neighborhood nodes gives the best results.

\paranofull{-- Number of epochs:} We ran our model and the simple MLP baseline for several epochs and plotted loss curves when validating their prediction capability in a small subset of the training data (using a 90:10 split). We observed that for more than 100 epochs there is no considerable change in the training/validation loss. Thus, in these experiments we train for 100 epochs. In the case of incremental training and since we have 10 relatedness graphs ($|\mathcal{RG}| = 10$) we train incrementally for 10 epochs per relatedness graph, leading to 100 epochs in total.

\paranofull{-- Dimension of embeddings:} We evaluated the effectiveness of our model for varying dimensions of node representations produced by the GraphSAGE model we use in the range of $\{32, 64, 128, 256, 512\}$. We found out that using embeddings of 256 dimensions provides with the best results.

\begin{table}[t!]
\centering
 \small
\resizebox{\columnwidth}{!}{
\begin{tabular}{l|c||c||c||c}
\toprule\bottomrule
\textbf{Benchmark} & \textbf{\# Silos} & \textbf{\# Datasets} & \textbf{\# Training Matches} & \textbf{\# Test Matches}  \\ \toprule\bottomrule
\textbf{NYC OpenData} & 10 & 290 & 5437 & 34537 \\ \hline
\textbf{LA OpenData} & 10 & 224 & 5913 & 34986
\\
  \toprule\bottomrule
\end{tabular}
}
\vspace{1mm}
\caption{Data silo matching benchmarks used for evaluation.}
 \label{tab:silo_config}
\vspace{-6mm}
\end{table}

For training we use the Adam optimizer \cite{kingma2014adam} with a learning rate of \texttt{0.01}, while we use a MLP of one hidden layer for both \method{} and the baseline.  Furthermore, our method is implemented in Python \texttt{3.7.4} and is openly available for experimenation\footnote{\url{https://github.com/delftdata/SiMa}}, while GraphSAGE was implemented using the Deep Graph Library \cite{wang2019deep} on top of PyTorch.\footnote{\url{https://pytorch.org}} Experiments for \method{} and the MLP baseline ran on an 8-core MacBook Pro, while for running COMA we set up a Linux machine with 128 AMD EPYC 7H12 2.60GHz cores.

\subsection{Effect of Optimizations}
\label{sec:exp_opt}
We assess the effectiveness of different negative sampling techniques and training schemes, as discussed in Section~\ref{sec:opt}. To this end, we run two sets of experiments: \emph{i}) using the incremental training scheme, we apply four variants of \method{}, where three are based on a different negative sampling strategy and one considers all negative edges without sampling, and \emph{ii}) using the best such variant, we compare \method's incremental training against training on the whole set of relatedness graphs we get from the data silos.

\para{Sampling Strategies} In Figure \ref{fig:ablation} we see precision-recall curves for \method's different variants when evaluated upon both data silo benchmarks. First, we validate the boost in effectiveness that sampling edges from each other domain per node, i.e. NS3, can bring. Particularly, we see a considerable increase in both precision and recall, since with NS3 every node receives negative edges that cover the spectrum of other domains present in the corresponding relatedness graph. Consequently, the false positive links that our method predicts decrease (i.e. precision increases), while the better representational quality of the embeddings produced by our encapsulated GraphSAGE model ensures fewer false negatives (i.e. recall increases).  Moreover, we see that using \method{} with NS3 can produce a higher precision for high recall values, especially in the case of LA OpenData.

On the other hand, the other two sampling techniques, NS1 and NS2, and the variant using all negative samples exhibit different results depending on each benchmark. In specific, we see that sampling edges per node, as specified by NS2, produces low effectiveness results in both data silo settings. Precision for high recall values is mediocre, due to the lack of diversity and completeness about the knowledge each node receives about other domains in the relatedness graph. Surprisingly, picking negative edges at random on the whole graph, as specified by NS1, seems to bring consistently better results than the NS2 variant, even if it does not guarantee that every node is covered by the negative edges sampled. However, NS1 may pick negative edges that are more informative, yet this cannot be guaranteed due to its randomness.

Finally, we observe that using all available negative edges without sampling brings inconsistent results. In Figure \ref{fig:ablation_ny} we see the variant using all negative edges during training performs better than employing NS1 and NS2, while it is very close to NS3. On the contrary, in Figure \ref{fig:ablation_la} \method{} with all negative edges results is worse than using NS1 and only slightly better than NS2. This behavior is to be expected, since not employing a dedicated sampling strategy that guarantees the quality and amount of negative edges included during  training (like NS3), means that the model risks overfitting.

\mybox{\textbf{Takeaways:} $i)$ among the different sampling strategies, sampling per domain -- NS3, yields the best results; $ii)$ removing negative sampling harms effectiveness.}

\para{Incremental Training} Here we want to verify whether incremental training has a substantial influence on the effectiveness of the training process. We observe that training on all relatedness graphs from the beginning can severely affect the effectiveness of our method, since our model overfits on the set of possible and negative samples it receives. In contrast, our incremental training scheme drastically helps our model to adapt to new examples and significantly improves its prediction correctness. Indeed, as seen on the right-hand side of Figure \ref{fig:ablation} by applying \method's model on every relatedness graph incrementally in the order of number of edges they store, we make sure that the learning process can leverage the novel information that each graph brings, i.e. novel examples of semantic types that were not introduced by the previous graphs. Moreover, incremental training ensures faster execution times, since in earlier epochs the model sees less training examples. Therefore, in the following experiments we configure \method{} to apply the incremental training scheme and use NS3 as the negative edge sampling strategy.

\mybox{\textbf{Takeaway:} \method{}'s incremental training scheme improves the effectiveness of \method{} as shown by the precision-recall curves, with higher precision for high recall values.}
\vspace{-3mm}
\begin{figure*}[t!]
    \centering
    \hspace{3mm}\begin{minipage}{\columnwidth}
            \includegraphics[width=.9\columnwidth]{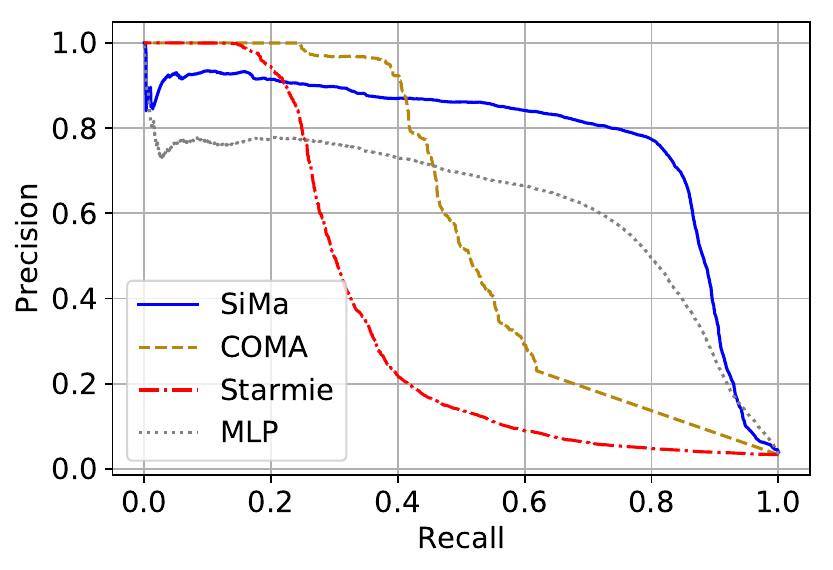}
            \vspace{-2mm}
            \subcaption{NYC OpenData}
            \label{fig:comparison_ny}
    \end{minipage}
    \begin{minipage}{\columnwidth}
             \includegraphics[width=.9\columnwidth]{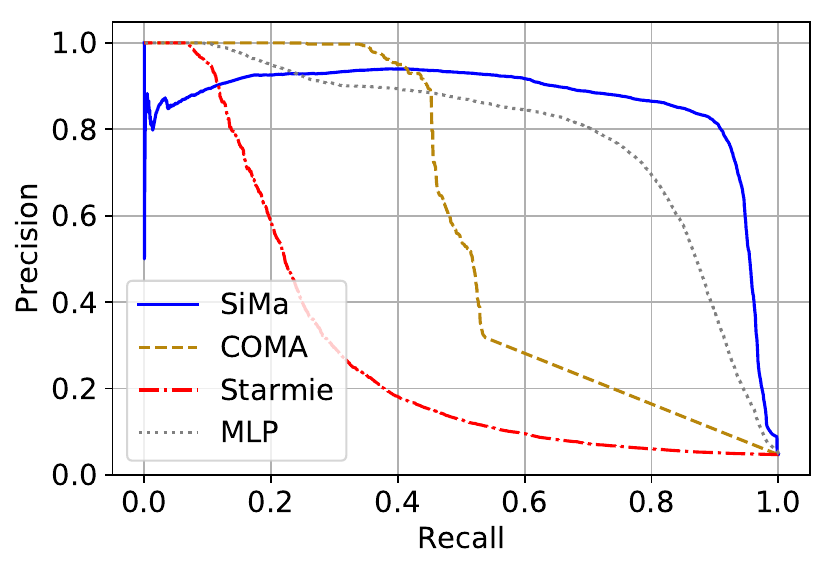}
             \vspace{-2mm}
             \subcaption{LA OpenData}
             \label{fig:comparison_la}
    \end{minipage}
        
    \vspace{-3mm}      
    \caption{Precision-Recall curves of \method{} and other methods.}
    \label{fig:comparison}
    \vspace{-3mm}
\end{figure*}

\subsection{\method{} comparison to other methods}
\label{sec:exp_comparison}
We compare \method{} with COMA, Starmie and the MLP baseline, to showcase the capability of our method to achieve better results in both effectiveness and efficiency. For a fair comparison with the MLP baseline we train it using the best negative sampling technique as found in \autoref{sec:exp_opt}, i.e., NS3, while we do so by employing the incremental training scheme. Below, we discuss the results.

\para{Effectiveness comparison} \autoref{fig:comparison} shows the comparison of \method{} against COMA \cite{do2002coma}, Starmie \cite{fan2022semantics} and the MLP baseline, in terms of effectiveness. First, we observe that \method{} learns significantly better how to disambiguate between positive and negative links, based on the knowledge that exists in each data silo, than the MLP baseline. For both data silo benchmarks we see that using only the initial column profiles with a simple MLP prediction model does not give good results. Indeed, existing matches between columns that are represented by profiles that are not similar,  cannot help a model. On the contrary, \method{} can learn the intrinsic graph characteristics that lead to a column relationship, by exploiting the message passing component of GNNs.

Surprisingly, even in the case where we could employ the state-of-the-art schema matching method of COMA for matching data silos, we observe that it would give inferior results compared to \method. In particular, \autoref{fig:comparison_ny} shows that COMA cannot keep a high precision for recall values above 0.4, which means that there is only a small fraction of matches that it can correctly predict. Similarly, in \autoref{fig:comparison_la} we see that COMA's precision  significantly drops for recall values above 0.5. On the contrary, \method{} in both cases can be highly precise even for recall values above 0.8. This is due to the fact that the similarity signals that COMA uses are oftentimes not sufficient to distinguish whether a pair of columns is a match or not; however, existing matches in the silos and the architecture of \method's  model enable our method to accurately sort out true negatives. Notably, our model outperforms COMA even if matching columns in the benchmarks we created have similar or exactly the same names, which is something that COMA takes advantage of. In a real world scenario, column names might not be human-understandable or could be missing, which would considerably decrease the effectiveness of COMA. \method{} is agnostic to column names, hence its performance is not affected by their existence/quality.

Finally, we observe that the contextualized column representations trained through BERT \cite{devlin2018bert} with Starmie produce results of low quality. In specific, we noticed that the false negative rate is significantly high when considering cosine similarity of Starmie embeddings between columns. Nonetheless, this result is expected: using context information to find column matches among datasets of different silos is not effective in our case, since most of the matches represent joins of columns that share no common context. In the original paper of Starmie \cite{fan2022semantics} such column representations are shown to be effective due to the nature of the problem that is targeted there: discovering unionable tables, requires a method that captures well the context of their columns.

\mybox{\textbf{Takeaways:} $i)$ \method{} exhibits consistently high effectiveness, whereas the competition falls short in precision for high recall values; $ii)$ embeddings computed through GNNs, have higher representational power than initial column features (MLP baseline); $iii)$ contextualized column representations are not suitable for matching columns across data silos.}

\begin{table}[tb]
\centering
 \small
%\resizebox{\columnwidth}{!}{
\begin{tabular}{l|c||c||c||c}
\toprule
\multicolumn{5}{c}{\textbf{Best F1 Scores}}\\
\bottomrule
\textbf{Benchmark} & \textbf{\method} & \textbf{COMA} & \textbf{Starmie} & \textbf{MLP} \\ \toprule
\bottomrule
\textbf{NYC OpenData} &  \textbf{0.787} &  0.564 & 0.384&  0.656\\ \hline
\textbf{LA OpenData} & \textbf{0.858}  & 0.600 &0.310  &0.736\\
  \toprule
  
  \multicolumn{5}{c}{\textbf{PR-AUC Scores}}\\
\bottomrule
\textbf{Benchmark} & \textbf{\method} & \textbf{COMA}& \textbf{Starmie}  & \textbf{MLP} \\ \toprule
\bottomrule
\textbf{NYC OpenData} & \textbf{0.774} &0.561& 0.358 &0.619\\ \hline
\textbf{LA OpenData} &\textbf{0.861}  & 0.578 &0.292 &0.761\\
  \bottomrule
\end{tabular}
%}
\vspace{1mm}
\caption{Effectiveness scores of \method{} and competition.}
 \label{tab:f1_auc}
\vspace{-6mm}
\end{table}
\begin{table}[t!]
\centering
 \small
%\resizebox{\columnwidth}{!}{
\begin{tabular}{l|c||c||c||c}
\toprule\bottomrule
\textbf{Benchmark} & \textbf{SiMa}  & \textbf{COMA} &  \textbf{Starmie}&\textbf{MLP}  \\ \toprule\bottomrule
\textbf{NYC OpenData} & \textbf{52} & 30900  & 73& 59 \\ \hline
\textbf{LA OpenData} & \textbf{51}  & 20100 & 61 & 54 \\
  \toprule\bottomrule
\end{tabular}
%}
\vspace{1mm}
\caption{Total execution times in minutes (CPU).}
 \label{tab:efficiency}
\vspace{-5mm}
\end{table}

\para{Efficiency comparison} In \autoref{tab:efficiency}, we see how \method{} compares with the other method in terms of efficiency measured in minutes. The total execution time for \method{} and the MLP baseline refers to the sum of dataset profiling, training and inference times.  

First, we observe that \method{} is considerably cheaper than the state-of-the-art traditional matching method COMA. Specifically, \method{} is more than two orders of magnitude faster. \method's runtime is dominated by the computation of profiles (roughly $80\%$ of total execution), hence in the case where these are pre-computed our method can give results in a small fraction of the time shown in the table. Additionally, we verify  that employing state-of-the-art schema matching methods, in this scale, might be infeasible: in real-world scenarios where datasets of multiple data silos with variable sizes need to be matched this can be prohibitively expensive. Specifically, COMA's syntactic similarity-based matching can be slow due to computations of various measures among instance sets of columns (e.g. TF-IDF), especially in the case where there are a lot of text values. 

On the other hand, we observe that using the initial profiles of the columns for training a simple prediction model with the MLP baseline not only is much less effective, but also exhibits slower training times. This is because the dimensionality of the initial profiles is much larger than the ones produced through the GraphSAGE model we employ  in \method. In addition, Starmie is slow when ran on CPU due to the computationally intensive training of the contextualized column representations through BERT, and the generation of positive and negative examples for its contrastive learning process.

\mybox{\textbf{Takeaway:} the complete pipeline of \method{} (profile computation, graph construction, training and inference) requires orders of magnitude less time and resources than the best-performing schema matching algorithm. This is due to the use of lower-dimension GNN embeddings for training our prediction model.}

\section{Conclusion}

In this paper, we introduced \method, a novel method for matching columns across disparate data silos, which uses an effective prediction model based on the representational power of GNNs. \method{} uses the knowledge about existing relationships among datasets in silos, in order to build a model that can capture potential links across them. Our experimental results show that \method{} can be more effective than state-of-the-art matching and column representation methods, while it is significantly faster and cheaper to employ. Moreover, we show that our optimization techniques significantly improve  the effectiveness of our method.

%\para{Future Work} \method{} makes the first step towards finding true matches across silos. A very important problem that we are focusing on at the moment is the one of automated data augmentation using matches, and building an exhaustive set of join paths among datasets. Those join paths need to be ranked according to different criteria. For instance, how well the resulting dataset of each join path helps in augmenting datasets used for building higher-accuracy ML models \cite{chepurko2020arda,zhao2022leva}.

\balance

\bibliographystyle{ACM-Reference-Format}
\bibliography{references}

\balance

\end{document}